\documentclass{aa}  
\usepackage{subfig}
\usepackage{graphicx}
\usepackage{txfonts}
\usepackage{natbib,twoopt}
\usepackage[breaklinks=true]{hyperref} 

\newcommand{\alfa}{$\alpha$}
\newcommand{\AF}{[\alfa/Fe]}
\newcommand{\Meta}{[Fe/H]} 

\newcommand{\T}{$T_{\rm eff}$}
\newcommand{\g}{log($g$)}

\newcommand{\Vmi}{$V_{\text{micro}}$}
\newcommand{\Vma}{$V_{\text{macro}}$}
\newcommand{\Vsini}{$V_{\text{sin $i$}}$}
\newcommand{\Nc}{$N_5$}
\newcommand{\Nt}{$N_3$}
\newcommand{\Nu}{$N_1$}

\bibpunct{(}{)}{;}{a}{}{,} 
\makeatletter
\newcommandtwoopt{\citeads}[3][][]{\href{http://adsabs.harvard.edu/abs/#3}%
        {\def\hyper@linkstart##1##2{}%
                \let\hyper@linkend\@empty\citealp[#1][#2]{#3}}}
\newcommandtwoopt{\citepads}[3][][]{\href{http://adsabs.harvard.edu/abs/#3}%
        {\def\hyper@linkstart##1##2{}%
                \let\hyper@linkend\@empty\citep[#1][#2]{#3}}}
\newcommandtwoopt{\citetads}[3][][]{\href{http://adsabs.harvard.edu/abs/#3}%
        {\def\hyper@linkstart##1##2{}%
                \let\hyper@linkend\@empty\citet[#1][#2]{#3}}}
\newcommandtwoopt{\citeyearads}[3][][]%
{\href{http://adsabs.harvard.edu/abs/#3}
        {\def\hyper@linkstart##1##2{}%
                \let\hyper@linkend\@empty\citeyear[#1][#2]{#3}}}
\makeatother

\begin{document}

\title{GSP-spec line list for the parametrisation of $Gaia$-RVS stellar spectra}


   \author{G. Contursi\thanks{Send offprint requests to Patrick de Laverny}
                        \and
          P. de Laverny
          \and 
          A. Recio-Blanco
          \and
          P. A. Palicio
          }

   \institute{Université Côte d'Azur, Observatoire de la Côte d'Azur, CNRS, Laboratoire Lagrange, Bd de l'Observatoire, CS 34229, 06304 Nice cedex 4, France\\
             }

   \date{Received ??; accepted ??}

 
\abstract
   {The $Gaia$ mission is a magnitude-limited whole-sky survey that collects an impressive quantity of astrometric, spectro-photometric and spectroscopic data. Among all the on-board instruments, the Radial Velocity Spectrometer (RVS) produces millions of spectra up to a magnitude of G$_{RVS} \sim 16$. For the brightest RVS targets, stellar atmospheric parameters and individual chemical abundances are automatically estimated by
 the Generalized Stellar Parametriser - spectroscopy group (GSP-Spec). These data will be published with the third $Gaia$ Data Release.}
   {Some major ingredients of the determination of these stellar parameters include 
   the atomic and molecular line lists that are adopted to compute reference synthetic spectra,
   on which the parametrisation methods rely. 
   We aim to build such a specific line list optimised for the analysis of 
   RVS late-type star spectra.}
   {Starting from the $Gaia$-ESO line lists, we first compared the observed and synthetic spectra of six well-known reference late-type stars in the wavelength range 
   covered by the RVS instrument. We then improved the quality of the atomic
   data for the transitions presenting the largest mismatches.}
   {The new line list is found to produce very high-quality synthetic
   spectra for the tested reference stars and has thus been adopted within GSP-Spec. 
   We note, however, that a couple of atomic line profiles, in particular
   the calcium infrared triplet lines, still show some deviations compared to the reference spectra,
   probably because of the adopted line-transfer assumptions 
   (local thermodynamical equilibrium, hydrostatic, and no chromosphere). Future works should focus on such
   lines and should extend the present work towards OBA and M-type stellar spectra.}
   {}


\maketitle
%

\section{Introduction}
Launched in December 2013, the ESA $Gaia$ mission is continuously surveying the sky to collect astrometric, spectro-photometric and spectroscopic data with an unprecedented precision for stars brighter than a given magnitude (down to G$\sim$20.7 and $\sim$16 for astrometry and spectroscopy, respectively). 
The data are then analysed by the Data Processing and Analysis Consortium (DPAC hereafter), 
which publishes successive versions of the $Gaia$ catalogue \citep[the last release 
being the early Data Release 3 catalogue; eDR3,][]{2020arXiv201201533G}. This eDR3 contains positions, parallaxes, proper motions
for about 1.5 billion stars, their magnitude in the $G$ band for around 1.8 billion stars, and radial velocities 
for 7.2 million stars.
The complete DR3 will be published in 2022. In particular, the spectroscopic part will contain a larger catalogue
of radial velocities. Stellar atmospheric parameters, chemical
abundances of several chemical species, and an analysis of a diffuse interstellar band (DIB) will be also published,
thanks to the parametrisation of the Radial Velocity Spectrometer data (RVS hereafter). 
This RVS instrument is described in detail in \citet{2018A&A...616A...5C}. The
collected spectra cover about 25~nm around the calcium triplet infrared lines 
($\sim$850~nm)
at a resolution close to 11,500. These spectra
are produced within the DPAC/Coordination Unit 6, which also derives the 
stellar radial velocities \citep{2018A&A...616A...6S}.
Then, the spectra of late-type stars are analysed by the Generalized Stellar Parametriser - spectroscopy (GSP-Spec) group \citep[see][and Recio-Blanco et al. 2022, in preparation]{GSPspec16} within the DPAC Coordination Unit 8 in
order to estimate the stellar effective temperature \T, the surface gravity \g, the global metallicity [M/H], and the abundance of $\alpha$-elements versus iron [$\alpha$/Fe].
In addition, individual abundances of several chemical species are also
derived. These stellar parameters and chemical abundances are automatically estimated 
thanks to different algorithms as the MATISSE and GAUGUIN methods \citep{2006MNRAS.370..141R, GSPspec16}. 
Finally, an estimate of the $Gaia$ DIB equivalent width is also performed within GSP-spec, following
a specific methodology presented in \citep{GDIB21}.
These parametrisation tools rely on grids of synthetic spectra that are used to train the algorithms and/or for references. 
Therefore, the accuracy of the derived stellar
parameters depends on the quality of these reference spectra. A special effort was thus
made within GSP-Spec to compute late-type star synthetic spectra as realistic as possible.
In particular, producing high-quality synthetic spectra requires a collection of complete
and high-quality atomic and molecular line data such as, for instance, line positions, oscillator 
strengths, excitation energies, and broadening parameters.

Important databases such as the Vienna Atomic Line Database (VALD hereafter, \citet{1995A&AS..112..525P}, \citet{2015PhyS...90e4005R}) or the National Institute of Standards and Technology Atomic Spectra Database (NIST, hereafter) provide a huge amount of line data to
synthetise stellar spectra. However, among the millions of lines that should be considered, only a small portion have precisely determined 
parameters. For instance, several lines observed in the solar spectrum still remain unidentified and/or are poorly synthetised.
Over the years, many efforts have been made to improve the quality of such line lists by reducing the uncertainty of the transition probability (e.g. \citet{2006JPCRD..35.1669F}, or the broadening parameters \citep[e.g.][]{Paul00}.
In this context and in order to improve the match between observed and synthetic spectra, several groups 
have already assembled high-quality
line data to build specific line lists devoted to late-type star studies.
In particular, in the wavelength range of the Ca~{\sc ii} triplet covered by the RVS instrument, an impressive
work has already been completed within the context of the $Gaia$ ESO Survey \cite[GES,][]{2021A&A...645A.106H} to build complete and high-quality atomic and molecular line lists.
Nevertheless, the high quality and the huge quantity of RVS spectra require a specific study
to check and improve (if necessary) the quality of the available line lists to infer good stellar parameters and/or chemical
abundances. 

The goal of the present article is to (i) quantify the quality of the available line data
for modelling late-type star spectra in the RVS spectral domain, and then (ii) present an improved version of it.
We first describe the adopted methodology to compare observed
and simulated spectra for well-defined reference stars in Sect.~\ref{Section:Obs-synt}. In Sect.~\ref{Sect:GES}, we quantify the quality of the GES line list and discuss the spectral transitions
that should be improved.
In Sect.~\ref{Sec.GL}, we outline how we improved the quality of the reference star
synthetic spectra leading to the line list
that has been adopted within GSP-Spec for the stellar parametrisation.
Finally, we conclude our work in Sect.~\ref{Sect:ccl}.

\section{Selection of the reference stars: Observed and computed spectra}\label{Section:Obs-synt}
In order to quantify the quality of a line list for the analysis of the $Gaia$-RVS spectra, 
we compare the spectra of six well-known reference stars against synthetic models both at high ($R \sim$ 100,000) and RVS spectral resolution ($R \sim$11,500). The details of the observed and synthetic spectra are explained in the following sub-sections.

\subsection{Reference stars and adopted atmospheric data}
\label{References}
Most of the RVS observed stars with accurate radial velocity estimates are of FGK  spectral types. We therefore selected reference stars belonging to these types in order to check the line list quality. These references must have very well-known atmospheric
parameters and individual chemical abundances together with available high-resolution ($R \gg 100,000$) and high 
signal-to-noise ratio (S/N $\gg$100) observed spectra
covering the RVS spectral domain. We first looked for such stars in the $Gaia$ benchmarks sample
of \citet{Benchmarks18} and finally selected six reference stars: $\mu$ Leo and Arcturus (two cool giant stars), 61~Cyg~A, $\epsilon$ Eri, the Sun (three cool dwarfs), and Procyon (hot dwarf) since they fulfil the above criteria. Their parameters are summarised hereafter.\\

For the Sun, 
we chose the observed spectrum of \citet{2011ApJS..195....6W}. This spectrum was collected at the McMath-Pierce telescope with the Fourier Tranfrom Spectrometer (see \citet{1975SoPh...41...43B}, \citet{1978fsoo.conf...33B}. It was collected over the integrated solar disc with a S/N exceeding several hundreds over the whole
optical domain (from 295.8 nm \ to 925.0 nm). Its spectral resolution is about 676,000 in our wavelength range.

The observed spectrum from Arcturus was taken from \citet{2000vnia.book.....H}. This spectrum covers  a wide wavelength range from 372.7 nm \ to 930.0 nm \ with a resolution of 150,000 and a S/N around 1,000. It was collected with the Coude Feed telescope on Kitt Peak using the spectrograph in the Echelle mode.

Procyon, 61~Cyg~A, $\epsilon$ Eri, and $\mu$ Leo spectra were taken from the PEPSI\footnote{\url{https://pepsi.aip.de/}} database.
PEPSI \citep{2015AN....336..324S,2018SPIE10702E..12S} is a high-resolution (up to 
$R \sim$ 270,000) echelle spectrograph for the 2x8.4 m Large Binoculary Telescope. It covers the optical/red domain from 383.0 to 907.0 nm and the collected 
spectra have typical S/Ns of several hundreds. \\
\begin{table*}[t]
        \caption{\label{RefStars} Atmospheric parameters and chemical abundances for our reference stars. 
        }
        \centering
        \begin{tabular}{lcccccc}
                \hline
                \hline
                Star &   \multicolumn{2}{c}{\it Cool giants} & \multicolumn{3}{c}{\it Cool dwarfs} & {\it Hot dwarf}\\
                     &   Arcturus& $\mu$ Leo & 61 Cyg A &  $\epsilon$ Eri & Sun & Procyon \\
                \hline
                \T~(K)          & 4286 & 4474  & 4374  & 5076  & 5771 &  6554 \\
                \g~(g in cm/s$^2$)              & 1.60 & 2.51  & 4.63  & 4.61  & 4.44 &  4.00 \\
                \Meta~(dex)     & -0.55 & 0.22  & -0.36 & -0.12 & 0.00&  -0.04 \\
                \AF~(dex)       & 0.22 & 0.12  & 0.02  & -0.01  & 0.00 & -0.07 \\
                \Vmi~(km/s)     & 1.95\tablefootmark{f}& 1.95\tablefootmark{f}  & 1.07\tablefootmark{e} & 1.14\tablefootmark{e} &0.87\tablefootmark{g} &1.69\tablefootmark{h}\\
                \Vma~(km/s)     & 4.30\tablefootmark{f} & 2.58\tablefootmark{f}  & 1.83\tablefootmark{g}   & 2.91\tablefootmark{g}  & 3.57\tablefootmark{g} & 4.60\tablefootmark{h} \\
                \Vsini~(km/s)    & 3.80\tablefootmark{f} & 5.06\tablefootmark{f}  & 0.70\tablefootmark{d}   & 2.40\tablefootmark{g} & 1.60\tablefootmark{g} & 2.80\tablefootmark{h}\\
                $\text{[C/Fe]}$~(dex)  & 0.43\tablefootmark{a} &  -0.18\tablefootmark{b} & 0.60\tablefootmark{c} & -0.06\tablefootmark{c} & 0.00 & -0.05\tablefootmark{c} \\
                $\text{[N/Fe]}$~(dex)    & 0.00 & 0.37\tablefootmark{b} & 0.00  & 0.00 & 0.00 & 0.00 \\
                $\text{[O/Fe]}$~(dex)    & 0.50\tablefootmark{a} &  -0.13\tablefootmark{b} & 0.47\tablefootmark{c} & -0.17\tablefootmark{c} & 0.00 & 0.14 \tablefootmark{c} \\
                $\text{[Ca/Fe]}$~(dex)  & 0.12 &  0.03 & -0.03 & 0.04 & 0.00 & 0.03 \\
                $\text{$^{12}$C/$^{13}$C}$  & 6.30\tablefootmark{i} &  20.0\tablefootmark{i} & 89.9 & 89.9 & 89.9 & 89.9 \\
                \hline
        \end{tabular}
        \tablefoot{The data come from \cite{Benchmarks18} or are assumed solar-scaled abundances except the following:
                \tablefoottext{a}{\citet{2011ApJ...743..135R},}
                \tablefoottext{b}{\citet{1990A&A...234..366G},}
                \tablefoottext{c}{\citet{2017AJ....153...21L},}
                \tablefoottext{d}{\citet{1984A&A...138..183B},}
                \tablefoottext{e}{\citet{2015A&A...582A..81J},}
                \tablefoottext{f}{\citet{2007A&A...475.1003H},}
                \tablefoottext{g}{\citet{2005ApJS..159..141V}, and}
                \tablefoottext{h}{\citet{2010MNRAS.405.1907B}.}
                \tablefoottext{i}{\citet{2013ApJ...765...16S}.}
                
        }
\end{table*}

These selected reference stars have well-known atmospheric parameters (\T, \g, \Meta, \AF)
and good data for their chemical individual abundances, microturlulent (\Vmi), macroturbulent (\Vma), and rotational (\Vsini) velocities. 
The adopted values of these parameters together with their associated reference are reported in Table~\ref{RefStars}. They have been collected as follows:

The atmospheric parameters were adopted from \citet{Benchmarks18}. We remind the reader that in a series of articles these authors (including several groups and different technics) determined the atmopsheric parameters and individual abundances of the Gaia benchmark stars for ten chemical species from high-resolution and S/N spectra. The chemical analysis was performed by adopting independent methods and the effective temperature and surface gravity  from \citet{2015A&A...582A..49H} and metallicity from \citet{2014A&A...564A.133J}.

\indent The broadening parameters (\Vmi, \Vma \ and \Vsini) were
taken from different literature sources as indicated in the notes of Tab.~\ref{RefStars}.
For the two selected cool giants (Arcturus and $\mu$ Leo), we adopted the parameters of \citet{2007A&A...475.1003H}.
We considered those of \citet{2005ApJS..159..141V} and \citet{2010MNRAS.405.1907B} for the Sun and Procyon, respectively.
Finally, for 61~Cyg~A and $\epsilon$ Eri, we adopted the \Vsini \ estimated by \citet{1984A&A...138..183B} and
\citet{2005ApJS..159..141V}, respectively; the \Vmi \ of \citet{2015A&A...582A..81J}; and the estimated \Vma \
from Eq.~1 of \citet{2005ApJS..159..141V} using our adopted \T.

\indent To compute accurate synthetic spectra, we also need good individual chemical abundances and, in particular
those of \alfa-elements
and iron-peak elements, which are the main contributors to the spectral lines in our wavelength range. 
Most chemical abundances (except a few exceptions for some stars, see below) are from \citet{Benchmarks18}, taking into account that their solar iron abundance is equal to -0.03~dex.
                We defined the \AF \ ratio as the mean of the Mg, Si, Ca and, Ti abundances given in this compilation. We also adopted their [Fe/H] values to derive the [X/Fe] abundances since [X/H] is provided.
Among the data given in Tab.~\ref{RefStars}, we explicitly
report the [Ca/Fe] abundances used in the spectra calculation because of the presence
of the important Ca~{\sc ii} IR triplet lines found in the RVS domain.
If not available in the above-cited article,
we considered individual abundances scaled to the adopted \Meta \ and \AF \ values.

Ideally, we also need individual abundances of carbon and nitrogen since CN is the most representative molecule in the RVS domain for the parameters of our reference stars (see Fig.~3 of \citet{2012A&A...544A.126D}. Similarly,
oxygen and titanium abundances are required since we notice a small contribution of TiO, especially for cool dwarf stars. Finally, FeH molecular lines may also marginally contribute to cool-star spectra. We remind the reader that Ti and Fe abundances are reported by \citet{Benchmarks18}. 

\indent For Arcturus, we adopted C, N, and O abundances from \citet{2011ApJ...743..135R}. These abundances
were derived using atmospheric parameters 
that are in complete agreement with those adopted in the present work. 
We note that our adopted C abundance for Arcturus was derived from high-excitation C I lines and differs by around 0.4 dex from other estimations based on molecular lines (\citet{2013ApJ...765...16S}).  However, we did verify that our adopted carbon abundance leads to a better fit of Arcturus in the RVS spectral range.

\indent For $\mu$~Leo, we adopted the C, N, and O abundances from \citet{1990A&A...234..366G}. The atmospheric
parameters adopted by these authors are again completely consistent with those of Table~\ref{RefStars} (\T \ differs by only 76~K
and \Meta \ by 0.09~dex), and these abundances can therefore be safely adopted.
We note that the C abundance of these stars was estimated from molecular features.
 
\indent For the three reference dwarf stars, 61~Cyg~A, $\epsilon$ Eri, and Procyon, we adopted carbon and oxygen abundances from \citet{2017AJ....153...21L}, although their adopted atmospheric parameters slightly differ between this study and the present work. Indeed, the effective temperatures reported by \citet{2017AJ....153...21L} are 107, 47, and 100~K higher, respectively, whereas differences in surface gravity are always lower than 0.05~dex and are thus completely compatible with ours. As discussed below in Sects.~\ref{Sec:dwarf} \& \ref{Sect:CygvsArc}, we point out that for 61~Cyg~A and $\epsilon$~Eri a
slightly lower carbon abundance (by about {-0.2~dex}) leads to a better agreement between the observed and simulated spectra. Such a proposed lower carbon abundance could result from the different adopted \T \ in \citet{2017AJ....153...21L}
and in the present work. 
Since no nitrogen abundances were found for these stars, we assumed that [N/H] is scaled to [M/H].

\indent Finally, when computing the synthetic spectra, we adopted the solar isotopic $^{12}$C/$^{13}$C ratio for the dwarfs, whereas we used the ratio derived by \citet{2013ApJ...765...16S} for the two cool giant reference stars.


\subsection{Computation of the reference stars synthetic spectra}
\label{Sect:synthetic}
In order to build the line list adopted for the spectral analysis conducted within GSP-Spec,
we adopted in the present work exactly the same tools for computing the synthetic spectra.
For that purpose, we used version 19.1.2 of the TURBOSPECTRUM\footnote{More 
recent versions, as the 19.1.3 that considered 
Stark line broadening were not available when this work was performed.} code \citep{2012ascl.soft05004P} that computes the continuous opacity and solves the radiative transfer equation in the lines for a given model atmosphere, an assumed chemical composition 
and lists of atomic and molecular transitions. Line
transitions are supposed to be formed under local thermodynamic equilibrium (LTE) and hydrostatic equilibrium
is also assumed in the stellar atmosphere.

We adopted 1D MARCS \citep{2008A&A...486..951G} model atmospheres\footnote{\url{https://marcs.astro.uu.se/}}. 
These models are fully compatible with TURBOSPECTRUM and are computed with two types of geometry: plane-parallel or spherical, depending
on the stellar surface gravity. Following \citet{Heiter06}, we chose spherical models for \g~$\le$ 3.5 (giant stars, g in cm/s$^2$) and plane-parallel geometry when \g~> 3.5 (dwarf stars).
For each reference star of Sub-sect.\ref{References}, we used the tool available on the MARCS website in order to interpolate a specific model atmopshere at the atmospheric parameters of Table~\ref{RefStars}.

On the other hand, the chemical abundances of each star reported in Table~\ref{RefStars} were considered in TURBOSPECTRUM. For the Sun and as adopted in the MARCS models, we used \citet{2007SSRv..130..105G}
chemical abundances and isotopic compositions.
The code also takes into account the microturbulence parameter (\Vmi), whereas 
macroturbulent
(\Vma) and rotational velocities (\Vsini) were considered afterwards by convolving
the computed spectra with the required broadening profiles. We note that we used a radial-tangent profile for the macroturbulence.
References for these velocity parameters are also given in Table \ref{RefStars}.

The last (and for our purpose most important) ingredients required to compute synthetic spectra are the atomic and molecular line lists. As a first try, we started this study by adopting the line lists provided by the 
line list group of the $Gaia$ ESO Survey \citet{2021A&A...645A.106H}.
For the atomic lines, they started with line data extracted from the VALD database\footnote{\url{http://vald.astro.uu.se}}
and then optimised the selected data for the analysis of FGK-type spectra by generally favouring experimental atomic data. 
This line list takes into account hyperfine structures (hfs), isotopic splitting, and up-to-date broadening due to collisions
with neutral hydrogen atoms.
For the molecular transitions, we considered the line lists from the following species (mostly from GES): 
$^{12}$CH and $^{13}$CH \citep{2014A&A...571A..47M}, 
$^{12}$C$^{12}$C \citep{2013JQSRT.124...11B}, 
$^{12}$C$^{14}$N \citep{2014ApJS..210...23B}, 
$^{12}$C$^{15}$N and 
$^{13}$C$^{14}$N \citep{2014ApJS..214...26S}, 
CaH and ZrO (including the isotopologues $^{90-92, 94, 96}$Zr, B. Plez, private communication), 
OH (T. Masseron, private communication), 
SiH (Kurucz, 2010), and
FeH \citep{2003ApJ...594..651D}.
We point out that, contrarily to GES, we adopted the most recent, complete, and accurate available line lists for VO \citep{2016MNRAS.463..771M}
and TiO \cite[][including its isotopologues $^{46-50}$TiO]{2019MNRAS.488.2836M}. For this TiO line list,
these authors indeed pointed out its better quality with respect to previous ones for simulating cool star spectra. The better quality of these TiO line data was recently confirmed by \citet{2020A&A...642A..77P} for cool stars in which these lines become dominant.
Of particular interest for RVS spectra, they found substantial improvement in the 840-850~nm region.
Good improvements are also found in 850-880~nm,\ but they noticed that new laboratory measurements with
smaller uncertainties in this region of the E-X 0-0 band could help to further improve the present line list quality.
Since no lines from $^{12}$C$^{13}$C, $^{13}$C$^{13}$C, NH, and $^{24-25-26}$MgH were found in the RVS domain,
these species were thus discarded.
In total, our line list is composed of millions of molecular lines and a couple of thousand atomic lines.
We point out that most of the adopted data are mainly composed by empirically or theoretical (and therefore not
measured) atomic/molecular data and the quality of some of them could therefore still be checked. 
We hereafter refer to this line list as the 'GES' one since most data come from \citet{2021A&A...645A.106H}.

Finally, we computed the reference stars' synthetic spectra with an initial wavelength step of 0.001~nm between 846.0~nm\ and 870.0~nm\ in order to cover the RVS spectral range. This led to synthetic spectra with 24,000  wavelength points ($wlp$, hereafter). We then convolved them to produce high-resolution ($R$ = 100,000) 
and RVS-like resolution ($R$ = 11,500) spectra, assuming a Gaussian profile to mimic the instrumental effect. The RVS-like
spectra have then been re-binned with 800 $wlp$ to satisfy the Nyquist-Shannon sampling criterion.

\section{Comparison between the observed and computed spectra for the GES \textbf{line list}}
\label{Sect:GES}
\begin{table*}[h!]
        \centering
        \caption{\label{chi2_11500} Quality-fit parameters (QFP) computed from the comparison between the synthetic and observed spectra at $R$ = 11,500 (800 $wlp$) for
                the GES line list, the complete GSP-Spec line list ($GL$), or rejecting the core of the calcium triplet lines ($GL_{\rm noCaT}$). See associated text in Sect.~3 for a definition of the QFP $\chi^2$, N$_1$, N$_3$, and N$_5$.}
        \begin{tabular}{lcccccc}
                \hline
                \hline
                Star &   \multicolumn{2}{c}{\it Cool giants} & \multicolumn{3}{c}{\it Cool dwarfs} & {\it Hot dwarf}\\
                & Arcturus & $\mu$ Leo  & 61 Cyg A & $\epsilon $ Eri & Sun & Procyon\\
                &GES/GES$_{\rm noCaT}$&GES/GES$_{\rm noCaT}$&GES/GES$_{\rm noCaT}$&GES/GES$_{\rm noCaT}$&GES/GES$_{\rm noCaT}$&GES/GES$_{\rm noCaT}$\\     
                
                \hline
                $\chi^2$       & 0.23/0.04 & 0.31/0.19 & 0.08/0.05 & 0.21/0.05& 0.07/0.03 & 0.17/0.04 \\
                N$_5$       & 17/0    & 18/7    & 0/0           & 11/0    & 4/0    & 6/0 \\
                N$_3$           & 42/0    & 59/46   & 1/0               & 10/3    & 7/1    & 43/4 \\
                N$_1$           & 210/123 & 390/286 & 306/196   & 273/146 & 96/50  & 176/84\\             
                \hline
        \end{tabular}
\end{table*}

\begin{table*}[h!]
        \centering
        \caption{\label{chi2_100000} Same as Tab.~\ref{chi2_11500}, but for $R$ = 100,000 (24,000 $wlp$).}
        \begin{tabular}{lcccccc}
                \hline
                \hline
                Star &   \multicolumn{2}{c}{\it Cool giants} & \multicolumn{3}{c}{\it Cool dwarfs} & {\it Hot dwarf}\\
                & Arcturus & $\mu$ Leo  & 61 Cyg A & $\epsilon $ Eri & Sun & Procyon\\
                &GES/GES$_{\rm noCaT}$&GES/GES$_{\rm noCaT}$&GES/GES$_{\rm noCaT}$&GES/GES$_{\rm noCaT}$&GES/GES$_{\rm noCaT}$&GES/GES$_{\rm noCaT}$\\     
                
                \hline
                $\chi^2$       & 10.94/3.45 & 25.94/18.86 & 4.71/3.09 & 10.67/3.23      &  4.34/2.00 & 7.92/2.43 \\
                N$_5$  & 812/220  &2711/2022   &99/63     &446/186   & 311/115 & 471/182 \\
                N$_3$  &1811/809 &3581/2724   &594/405   &723/413  &511/435  & 1585/305 \\
                N$_1$  &5816/3170 &9079/6694 &9392/6385 &8972/5367&3854/1960& 5399/2946\\
                \hline
        \end{tabular}
\end{table*}

%

We now compare the observed and simulated spectra at high and low spectral resolutions ($R$=100,000 and 11,500) over the spectral domain [846.0 - 870.0~nm] with 24,000 and 800 $wlp$, respectively. Choosing these two resolutions 
allowed us to 
(i) quantify the consistency between the computed and observed spectra for
RVS-like data, (ii)
more easily detect the problematic lines since, thanks to the high resolution, blends are more
easily identified, and (iii) improve the line list quality by calibrating identified problematic lines if necessary (see next section).

In order to quantify the possible mismatches between observed and computed spectra,
they were compared in the vacuum with $Gaia$ spectra for consistency. We point out that the line list presented above contains air wavelengths and, thus, the spectra are computed in the air.
The air-to-vacuum conversion of the fluxes was then done afterwards thanks to the relation given in \citet{1994Metro..31..315B}. To quantify the differences,
we introduced the following quality fit parameters (QFP):

\indent First, the flux difference is quantified by $\chi^{2}$ = $\sum_{i = 0}^{N_{\mathrm{wlp}}} (\text{$M$[i]} - \text{$O$[i]})^2$, with
        $M$ and $O$ being the synthetic and observed spectra, respectively. $N_{\mathrm{wlp}}$, the number of $wlp$, is equal to 24,000 or 800, depending on the spectral resolution. 
        
        Then \textit{N$_1$, N$_3$, N$_5$} are the number of $wlp$ for which a difference between the synthetic and observed normalised fluxes are between 1\% and 3\%,  between 3\% and 5\%, and higher than 5\%, respectively. 

Tables \ref{chi2_11500} \& \ref{chi2_100000}
present these QFP for the six selected reference stars
computed at R = 11,500 and 100,000, respectively.
In both tables and for each star, the first column refers to the QFP for the synthetic spectra computed with the GES 
line list presented above in Sect.\ref{Sect:synthetic}. In order to more easily identify possible mismatches between observed and computed spectra, we calculated these QFP with and without considering the calcium triplet lines (GES$_{\rm noCaT}$ in the second columns of both tables) since accurately simulating their core
could be problematic because of NLTE and/or chromospheric effects. 
For this GES$_{\rm noCaT}$ case, we excluded a window of $\sim$2~nm width centred at the core of the three  Ca~{\sc ii} lines. The discarded wavelength ranges are (in vacuum and nm): [849.43-851.03], [853.73-855.73], and [865.74-867.74].

In the present section, we mostly focus on the results presented in Table \ref{chi2_11500} for the $Gaia$-RVS spectral resolution but the same tendency is also seen at higher resolution (Table \ref{chi2_100000}). 
First, we discuss the impact of the metallicity on the fit quality for the two reference cool giant stars in Sect.~3.1.
Then, we inspect the importance of the temperature on the QFP by examining the dwarf spectra in Sect.~3.2.
Finally, we examine the effect of the surface gravity on the fit quality by comparing spectra
of dwarfs and giants with close \T \ (Sect.~3.3).

\begin{figure}[t!]
        \centering
        \subfloat{
                \label{fig:CN-TiO-giant}
                
                \includegraphics[width=\hsize]{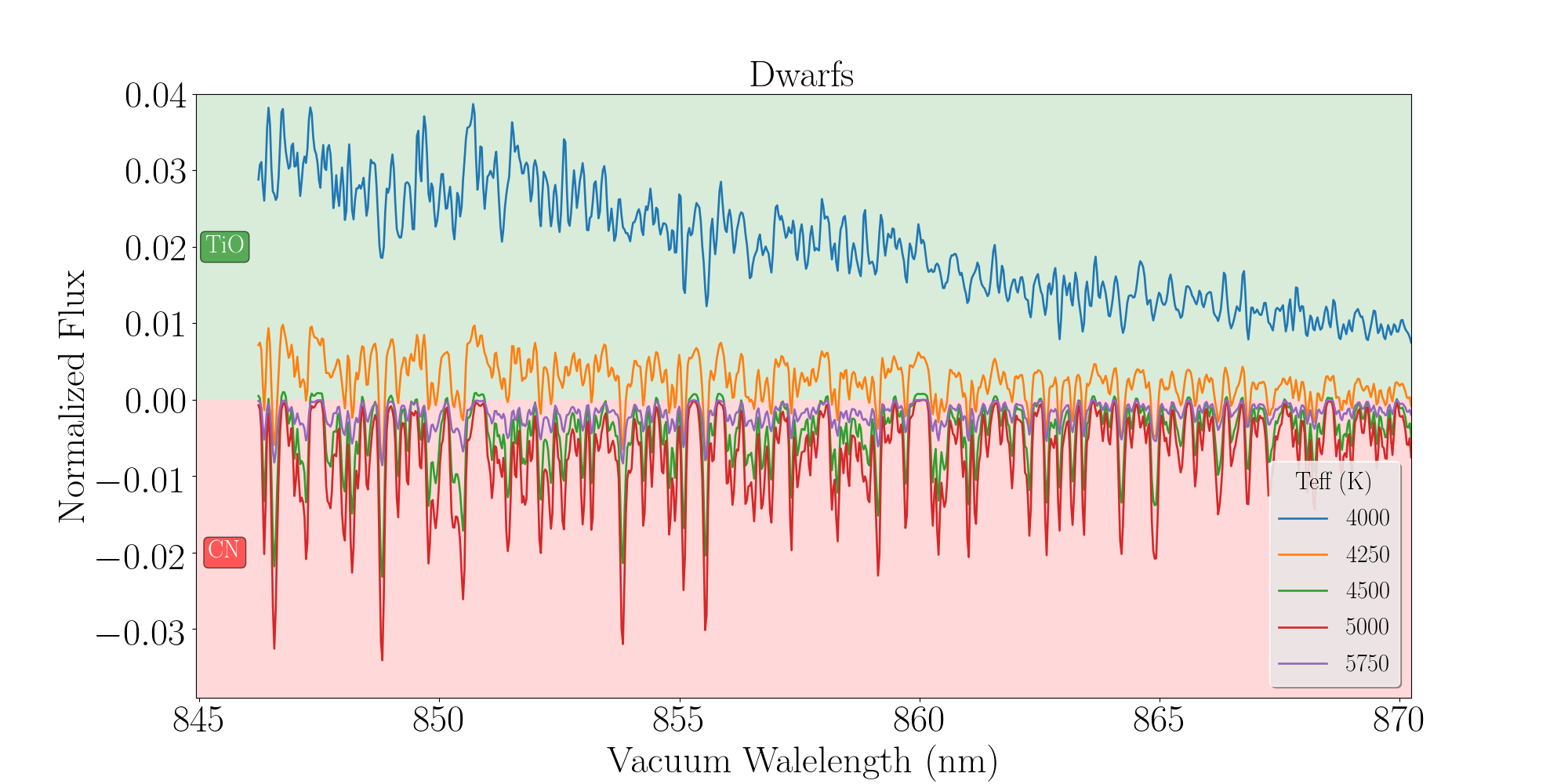} 
        }\\
        \subfloat{
                \label{fig:CN-TiO-dwarf}
                
                \includegraphics[width=\hsize]{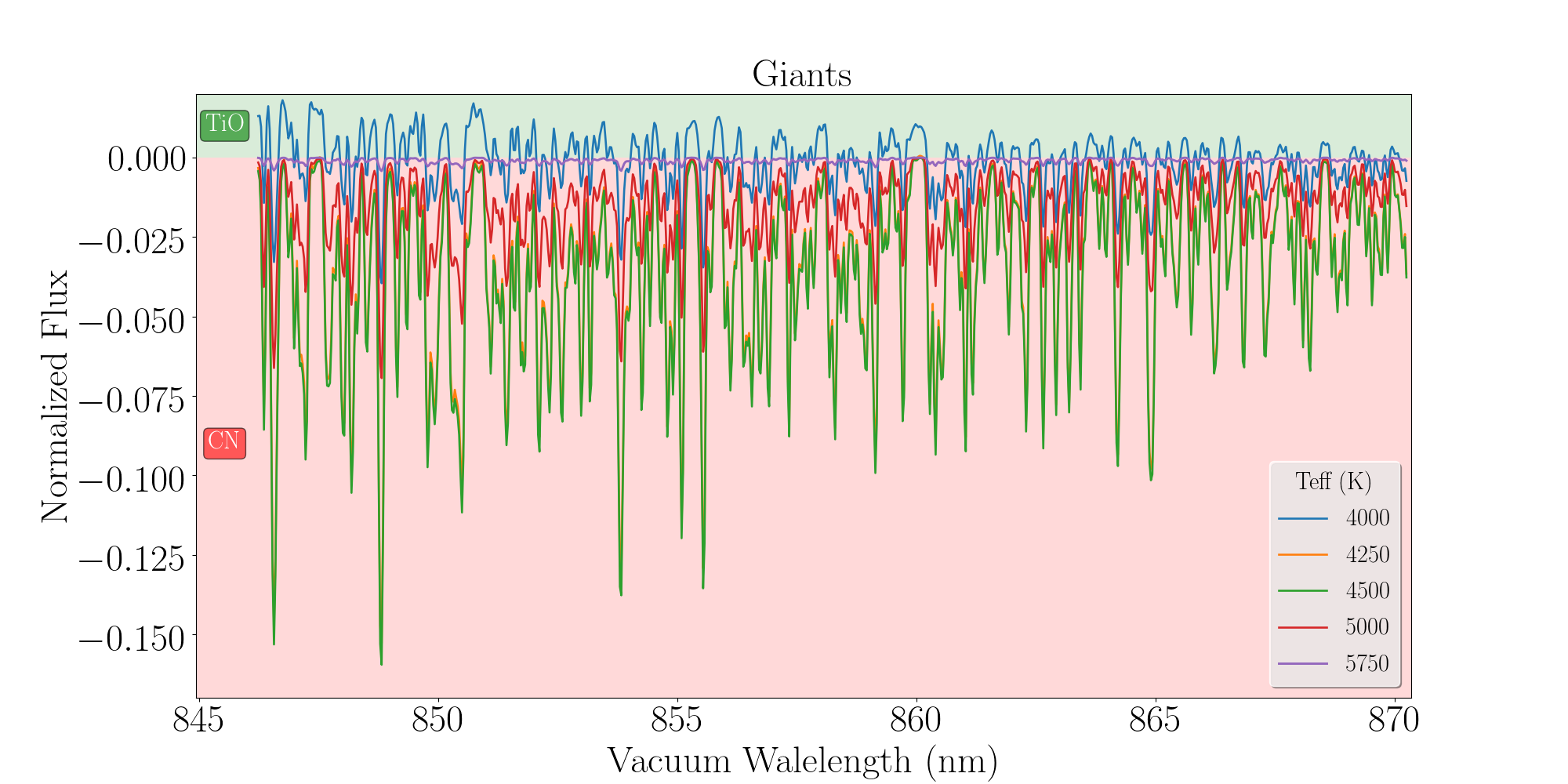} 
        }\\
        \caption{
        Differences between synthetic spectra (normalised flux) computed with only CN and TiO lines for five values of the effective temperature. The spectra were computed adopting \Meta \ = \AF \ = 0.0~dex with $R$=11,500 and 800 $wlp$ for typical dwarf (\g=4.5, \Vmi=1.0~km/s) and giant stars
        (\g=1.5, \Vmi=2.0~km/s), top and bottom panels, respectively.
        Green and red background colours refer to dominant contributions of TiO and CN, respectively.} 
        \label{fig:CN-TiO}
\end{figure}

\subsection{Cool giant stars}
First of all, we notice that among the two cool giants, Arcturus
exhibits the best fit both at high and low spectral resolution. 
For example, the $\chi^2$ and the number of $wlp$ with a difference between 3\% and 5\% ($N_3$) vary by more than a factor of two at high resolution between these two giant stars. 
The reason is that, although their rather similar \T \ and \g, Arcturus' spectrum exhibits far fewer 
lines than 
$\mu$~Leo because it is more metal poor ($\mu$~Leo being $\sim$0.75~dex more metal rich, 
see Tab.~\ref{RefStars}). 
It is therefore much easier to produce a more realistic spectrum and, hence, 
better QFP for Arcturus.

Moreover, for both stars (but this is also true for any other reference star), the large difference between \Nu \ and \Nt \ (or \Nc) is mostly dominated by molecular transitions that are predominantly fainter.
We point out that in these cool giant stars, the CN molecule is the dominant
species (TiO lines beginning to be visible only for \T$\la$4,100~K). We also checked that, for any cool stars, atomic lines mostly contribute to the major mismatches (\Nt \ and \Nc).
Therefore, the huge number of molecular transitions that appear
in the metal-rich spectrum of $\mu$~Leo explains its much larger \Nu \ with respect to other reference stars at low resolution.

Furthermore, by looking for each star at the column GES$_{\rm noCaT}$ of Tab.~\ref{chi2_11500}, it is remarkable that the calcium triplet lines considerably impact the fit quality. For instance, if we do not consider this triplet for Arcturus, the largest mismatches in flux 
(\Nc \ and \Nt) completely disappear at the RVS resolution, although smaller differences (\Nu) are still present but they are less numerous.
However, there are still some large discrepancies at higher resolution for Arcturus, mainly caused by atomic lines that require some 
improvements (see Sect.~\ref{Sec.GL}). Similar conclusions are obtained for the $\mu$~Leo spectrum when removing these
calcium lines.

\subsection{Dwarf stars}
\label{Sec:dwarf}
First, when comparing the three coolest dwarf stars (61~Cyg~A, $\epsilon$~Eri, and the Sun), we notice that there is a very good agreement between observed and synthetic spectra at the RVS resolution both in terms of \Nc \ and \Nt,
particularly if one disregards the calcium triplet lines. However, smaller flux differences (\Nu) are still numerous and are again dominated mostly by molecular lines. 

Then, since the surface gravities are similar and the metallicities almost do not differ among these dwarfs,
the main parameter that affects their fit quality is the effective temperature that varies by about 1,400~K. In particular, \T \ plays a crucial role since it governs the presence or absence of both atomic and 
(most importantly for \Nu) molecular lines. Regarding these molecular transitions, Fig.~\ref{fig:CN-TiO} (top panel) shows the difference between a spectrum calculated with only CN and TiO lines (including all their respective isotopes) at the low effective temperatures for which molecular lines 
are important in the stellar spectra (although still present but very weak at $\sim$ 5,750~K). We remind the reader that these two molecules are the most important ones for the selected reference stars and spectral domain. 
These dwarf spectra differences were calculated assuming 
\Meta =\AF = 0.0~dex, \g = 4.5, and \Vmi = 1.0 km/s at the RVS resolution. No broadening due to \Vma \ and \Vsini~ were considered.
When the difference between the CN and TiO contributions to the spectra is positive (negative), this means that the TiO lines contribute more (less) than CN ones (green and red backgrounds, respectively).
First, we point out that for \T $\la$4000~K and \T $\ga$4,300~K, these two molecules almost do not coexist
in dwarf spectra: TiO being present/absent (and CN absent/present) for these two temperature regimes, respectively. Moreover, it is noticeable that below $\sim$4,000~K, TiO is by far the dominant absorbing species. 
On the contrary, for larger \T, the contribution of CN becomes dominant and reaches a maximum around $\sim$5,000~K, corresponding to the effective temperature of $\epsilon$~Eri. This is the main reason why this star has the worst fit with respect to the two other cool dwarfs: far more molecular (CN) lines are present in its spectrum.
All of this can be explained by (i) the smaller dissociation energy of TiO with respect to CN, favouring the formation of CN at effective temperature hotter than $\sim$4,300~K; and (ii) at lower temperatures, the formation
of TiO is favoured since most carbon atoms are blocked onto CO molecules and, with oxygen being more abundant than 
carbon, almost no CN molecules can be formed. 
Finally, for even larger \T \ (becoming closer to the solar value), the contribution of the CN molecule decreases 
and becomes negligible since the stellar atmopshere becomes too hot to favour molecular formation. 
Nevertheless, one also notices that, for $\epsilon$~Eri, a better fit 
of the CN lines (and an improvement of all its
QFP) would be obtained by adopting a slightly lower carbon abundance (about -0.2~dex).
Decreasing the nitrogen abundance does not improve the global fit as well. 
Such a proposed lower carbon abundance would result from the slightly hotter \T \ (with respect to us) adopted in \citet{2017AJ....153...21L} when deriving their C abundance in 61~Cyg~A.


However, one can also see in both tables that \Nc \ and \Nt \ are better for 61~Cyg~A than for the hotter Sun. 
This is interpreted by (i) the larger metallicity of the Sun leading to 
the formation of more lines in its spectrum although its larger \T, \
and (ii) the better fit of the calcium IR triplet lines (including the wings)
in 61~Cyg~A. By looking at the GES$_{\rm noCaT}$ column of Tables \ref{chi2_11500} \& \ref{chi2_100000}, it is noticeable that, without the calcium triplet lines, all the QFP (except \Nt \ and \Nc \ at high resolution) become smaller for the Sun than for 61~Cyg~A. One can also see that the atomic lines in the solar spectrum are rather well synthesised (very good \Nc \ and \Nt \ at 11,500). This probably results from the fact that atomic line data are predominantly checked and/or optimised for this very well-studied  reference star.

Finally, it is well known that the effective temperature also acts on the number of atomic lines in stellar spectra:
a hot spectrum exhibits fewer atomic lines than a cooler one at constant metallicity. 
It is therefore logical that a good fit of hot star spectra should be easier to achieve with respect to cooler ones because of the difficulty to obtain accurate atomic and molecular line data.
This is confirmed by the QFP reported in Tables~\ref{chi2_11500} \& \ref{chi2_100000}, where it is noticeable that Procyon shows a better fit than $\epsilon$ Eri (neglecting the calcium triplet lines).
If one neglects these three calcium lines, the fit of Procyon becomes as good as the one of 61~Cyg~A and the Sun.
It is, however, slightly worse than in the Sun since some atomic lines predominantly formed in the hottest stellar atmosphere
(particularly seen at high resolution in \Nt \ and \Nc) still need to be improved.

\subsection{Comparing dwarfs and giants of similar effective temperatures}
\label{Sect:CygvsArc}
We now examine the effects of the stellar surface gravities on the fit quality by comparing
the QFP for 61~Cyg~A and Arcturus, which differ
by about 3~dex in \g \ but have almost similar \T \ and \Meta. Looking at the first column for both stars in Tables~\ref{chi2_11500} \& \ref{chi2_100000}, we first notice that the largest mismatches expressed through \Nt \ and \Nc \ are much larger in Arcturus than in 61~Cyg~A. Comparing for these two stars the second column of these tables, we can see that these large differences are mainly caused by the rather bad fit of the calcium triplet lines. Hence, removing them for the estimation of the QFP produces at low resolution (especially for these two stars) a fit of very good quality, except for the smallest flux differences (\Nu). These smallest mismatches are more numerous for 61~Cyg~A than for Arcturus. By comparing the top (dwarfs) and bottom (giants) panels of Fig.~\ref{fig:CN-TiO}, one can see that, at the studied \T \ ($\sim$4,300~K), the contribution of CN is much stronger in giants than in dwarfs (TiO being  almost unformed in both stars). We note that, in the bottom panel, we adopted \g =1.5 and \Vma =2.0~km/s (the other parameters being similar to those adopted for dwarfs in the top panel). However, although fainter in 61~Cyg~A, we checked
that its CN lines are not as well fitted as in Arcturus and produce its higher \Nu. This
could again be solved by adopting a lower carbon abundance by about $\sim$0.2~dex in 61~Cyg~A, as we already proposed for $\epsilon$~Eri.
Again, this proposed smaller [C/Fe] could result from the hotter \T \ adopted by \citet{2017AJ....153...21L}. 
Slightly decreasing the carbon abundance in this star
leads to a much better fit and a two-times-lower N1 (thus becoming smaller than in Arcturus).
Finally, we note that, apart from the calcium triplet lines, the majority of the largest mismatches between the observed and simulated spectra (especially at high resolution) for these two cool stars are caused by atomic lines whose fit could be improved. 

\section{The GSP-Spec line list}
\label{Sec.GL}

\begin{table*}[h!]
        \centering
        \caption{\label{GL_11500} Quality-fit parameters computed from the comparison between the synthetic and observed spectra at $R$ = 11,500 (800 $wlp$) for
                the complete GSP-Spec line list ($GL$) and rejecting the core of the Calcium triplet lines ($GL_{\rm noCaT}$).}
        \begin{tabular}{lcccccc}
                \hline
                \hline
                Star &   \multicolumn{2}{c}{\it Cool giants} & \multicolumn{3}{c}{\it Cool dwarfs} & {\it Hot dwarf}\\
                & Arcturus & $\mu$ Leo  & 61 Cyg A & $\epsilon $ Eri & Sun & Procyon\\
                &$GL/GL_{\rm noCaT}$&$GL/GL_{\rm noCaT}$&$GL/GL_{\rm noCaT}$&$GL/GL_{\rm noCaT}$&$GL/GL_{\rm noCaT}$&$GL/GL_{\rm noCaT}$\\       
                
                \hline
                $\chi^2$       & 0.20/0.03 & 0.26/0.14 & 0.08/0.06 & 0.17/0.04& 0.04/0.02 & 0.14/ 0.03 \\
                N$_5$    & 15/0    & 14/3     & 0/0 & 10/0 & 4/0 & 6/0 \\
                N$_3$    & 32/0           & 43/29    & 0/0 & 5/1 & 2/0& 22/0 \\
                N$_1$    & 166/82 & 358/285 & 313/222& 254/128& 32/21 & 160/54 \\
                \hline
        \end{tabular}
\end{table*}
        Although the global agreement between the observed and simulated spectra appears already rather good, there are still some 
        important local disagreements revealed by poorly synthesised spectral lines. 
        Apart from the calcium triplet lines, most of the largest spectra differences (\Nt \ and \Nc) for the reference stars are 
        caused by atomic transitions. 
        To improve this situation, we therefore decided to build a new atomic line list, starting from the one presented in Sect.~\ref{Sect:synthetic}. We remind the reader that we only focus on the atomic transitions since they play a crucial role in the atmospheric parameter and abundance derivations performed within GSP-Spec.
        This new line list was created by (i) identifying the atomic transitions (and associated possible blends) causing the largest 
        mismatches between the observed and simulated high-resolution spectra,  and then (ii) calibrating the atomic data 
        of these identified lines. This new list is called the GSP-Spec line list hereafter ($GL$).
        
\subsection{Astrophysical calibration of some atomic lines}
        
        We started by identifying
        the atomic transitions leading to the highest mismatches at both high and low resolution thanks to the \Nt~and \Nc \ QFP. Among the detected lines with uncertainties in the line positions, we only
        had to slightly correct 
        the wavelength of the multiplet~6 of S~{\sc i}. We adopted
        869.5524, 869.6319, and 869.7014~nm in vacuum (869.3137, 869.3931, and 869.4626~nm in air), in perfect agreement with \citet{Wiese69}.
 
        Then, for all the other identified lines, it appeared necessary to only correct their
        oscillator strength ($gf$) without changing any other of their line data such as their
        broadening parameters. These $gf$ were first 'astrophysically' calibrated to 
        improve the fit quality (visual inspection of the fit and check of the QFP) and match the solar spectrum as well as possible. 
        We checked that this calibration also improved the Arcturus and Procyon fits, these being the best known 
        stars among cool giants and hot dwarfs. For this calibration,
        we always assumed the chemical abundances
        reported in Tab.~\ref{RefStars}. Finally, as is shown 
        Sect.~\ref{Section:GL_quality}, these modified oscillator strengths improve
        the fit quality of all of our other reference stars. 

\begin{figure}[t!]
        \centering
        \resizebox{10.cm}{6.5cm}{\includegraphics{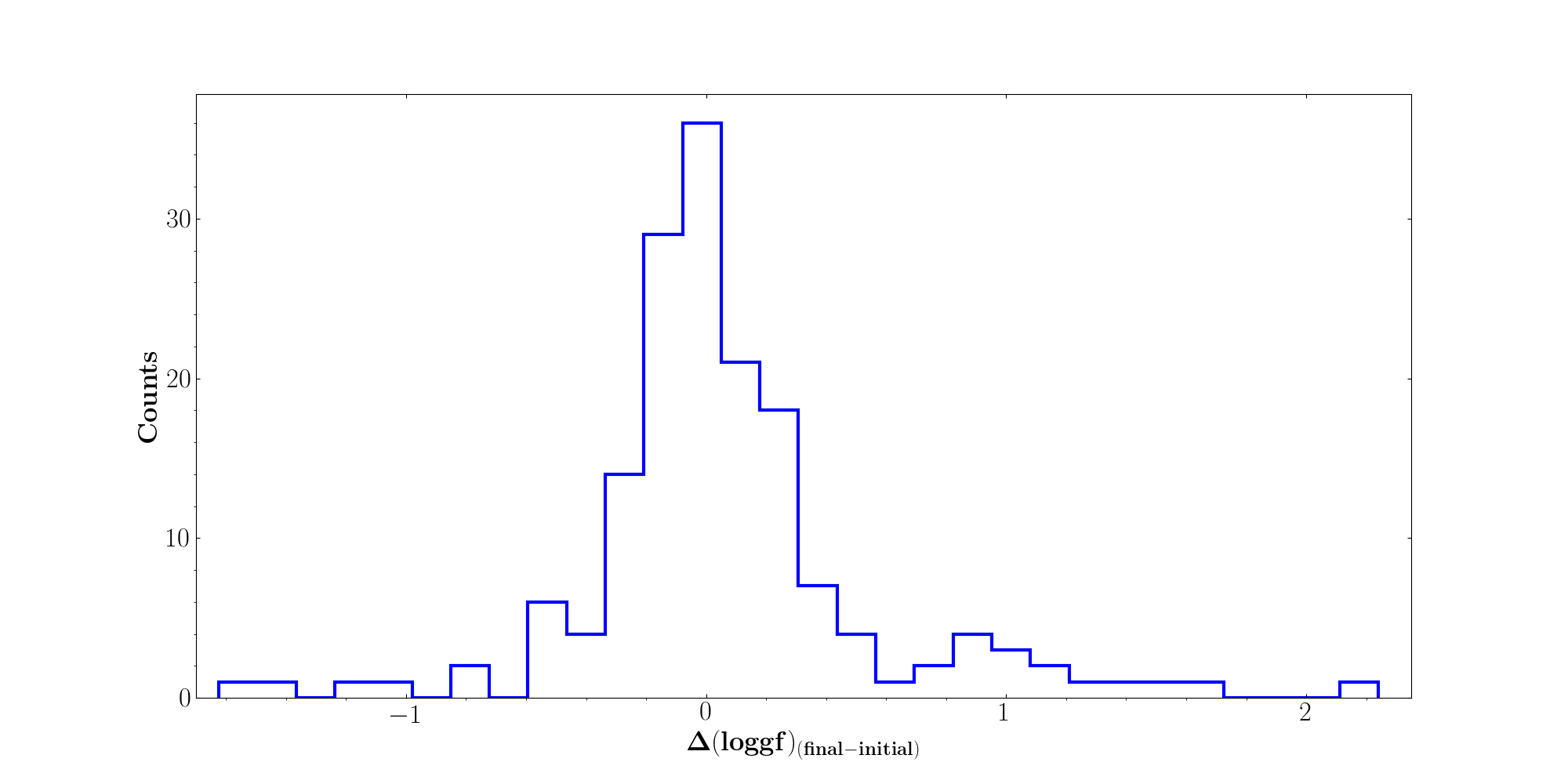}}
        \caption{Distribution of the differences between the initial and adopted 
        oscillator strengths for the astrophysically calibrated lines of the GSP-Spec line list.}
        \label{Fig:gf}
\end{figure}

        In total, we calibrated about 
        170 atomic oscillator strengths belonging to about ten different atomic species 
        (whatever their isotope and ionisation state) among the $\sim$70 chemical elements 
        present in the line list. All these calibrated transitions are listed in Table \ref{Lines}. The distribution of the differences between the initial and calibrated oscillator strengths is shown in Fig.~\ref{Fig:gf} (excluding the potential Fe~{\sc ii} line discussed hereafter because of its unknown real nature).
        The mean change in log($gf$) is 0.30~dex and 50\%, 80\%, 95\% 
        of the lines were corrected by less than 0.15, 0.50, and 1.10~dex, respectively. 
        Moreover, no systematic changes for a given species are noticed.
        These corrections lead to changes of up to a few percent in the flux of some lines 
        at the RVS resolution. Regarding the three huge calcium triplet lines that were causing
        the largest differences between observed and synthetic spectra, we 
        slightly decreased their oscillator strengths by 0.024/0.037/0.027~dex, favouring   
        the fit quality of the calcium wings since they contribute to several blends. 
        This led to a well-improved wing fit,
        although it was not perfect in all the reference stars since different
        behaviours in the line profiles are seen among these stars. 
        We note, however, that it was impossible to simultaneously well
        fit both the core and the wings of these calcium lines with the adopted physical assumptions
        since these cores are probably subject to NLTE and/or chromospheric effects. 
        Perfectly fitting both the cores and the wings of these CaT lines 
        would indeed require a more realistic treatment of the radiative transfer in these 
        stellar atmospheres, which is beyond the scope of the present work. 
        
       Finally, we remarked a line present in the observed spectrum of Procyon and also visible although fainter in the solar spectrum (with a normalised flux of 0.95\% and 0.97\%, at the RVS resolution, respectively).        By looking at spectra of stars hotter than Procyon, we verified that this line becomes stronger for hotter \T, 
       suggesting that it belongs to an ionised element and/or has a high excitation energy.
       This line is found in vacuum around 858.794~nm ($\sim$858.558~nm in air) and is absent from our synthetic spectra
       whatever the effective temperature is.
       This unknown line is therefore missing in the GES line list or has incorrect atomic parameters. 
       In order to identify it, we searched the VALD and NIST databases  as well as in Kurucz website \footnote{\url{http://kurucz.harvard.edu/linelists/}}for all lines between 858.500 and 858.600 nm (in air).
       First, we point out that \citet{1966sst..book.....M} identified this line as S~{\sc i,} but since no such sulfur line was reported by line databases, we discarded this proposition.  
       Then, the most likely candidates found were the Fe~{\sc ii} (858.544 nm in air; 858.779 nm in vacuum) and Sr~{\sc iii} (858.552 in air; 858.788 nm in vacuum) lines. The excitation energy of the latter (33 ev) seems to be too high to be observed in stars of temperature smaller than 8000~K. The other candidate lines being too distant in wavelength and/or too weak, we therefore considered this line 
       as corresponding to the Fe~{\sc ii} transition that was already present in the original
       GES list but with a far too weak oscillator strength. 
       In order to well fit our hot reference star spectra, we had to calibrate its wavelength 
       and $gf$, which was increased by 8.786~dex. We also
       note that this line seems to be blended by another unidentified line (around 858.850 nm in vacuum) that appears much weaker 
       in Procyon and solar spectra. 
       Nevertheless, this tentative Fe~{\sc ii} line was adopted in our GSP-Spec line list and well improves
        the fit quality of the hot star spectra. Its Fe~{\sc ii}-nature should be confirmed later by comparison with the
        abundances derived from other independent iron lines when analysing hot star spectra within GSP-Spec.
        
        \begin{table*}[h!]
                \centering
                \caption{\label{Lines} Calibrated atomic lines. Column 1 indicates the element and its ionisation state; Cols.~2 \&~3: air and vacuum wavelengths; Col.~4: lower excitation potential which comes from GES line list; 
                Col.~5: log(\textit{gf}) from the GES line list, whereas Col.~6 provides our calibrated value. The * close to the air wavelengths indicate that they were slightly modified compared to those given in GES. The spectral lines are ordered by increasing atomic number and increasing wavelength for each atomic species. The full table is available in electronic form at CDS. }
                \begin{tabular}{lccccc}
                        \hline
                        \hline
                        Element & $\lambda_{air}$ & $\lambda_{vac}$ & E  & log(\textit{gf})$_{GES}$ & log(\textit{gf})$_{RVS}$ \\
                         & (nm) & (nm) & (eV) &  &  \\
                         \hline
                         Na I & 864.8931 & 865.2307 & 3.191 & -1.997 & -2.040 \\
                         Mg I & 860.9727 & 861.2092 & 6.118 & -2.810 & -2.300 \\
                         Si I & 846.1482 & 846.3807 & 5.964 & -2.757 & -2.400 \\
                         Si I & 849.2077 & 849.4410 & 5.863 & -2.742 & -2.100 \\
                         Si I & 850.1544 & 850.3880 & 5.871 & -0.817 & -1.260 \\
                          . & . & . & . & . & . \\
                          . & . & . & . & . & . \\

                        \hline
                \end{tabular}
        \end{table*}

        
\subsection{Quality of the GSP-Spec line list}
\label{Section:GL_quality}

        To quantify the quality of the new fits between the observed and simulated spectra 
        adopting the improved GSP-Spec line list, 
        Tab.~\ref{GL_11500} presents the new computed QFP at the RVS resolution. It is built as Tab.~\ref{chi2_11500}, to which it should be compared. The first column for each reference star ('$G$L') refers to the QFP corresponding to the final adopted line list (the spectra being
        computed exactly as those described in Sect.~\ref{Sect:synthetic}). 
        The second column of each reference star ($G$L$_{\rm noCaT}$)
        shows the QFP by disregarding 
        $\sim$2 nm at the core of the three Ca~{\sc ii} lines as previously done.
        
        First, it can be clearly seen that a global improvement of the fit quality of any reference star is obtained when adopting the GSP-Spec line list: the $\chi^2$ are smaller
        and most of the mismatches previously reported have indeed disappeared.
        For instance, if one disregards the CaT lines, \Nc \ and \Nt \ become always null,
        except for a few $wlp$ in $\mu$~Leo caused by molecular lines
        and only one $wlp$ in $\epsilon$~Eri (a Si~{\sc i} line not well fitted only in this star,
        suggesting a possible different Si abundance than the one adopted). 
        A gain of about a factor of two in \Nt \ is obtained for $\epsilon$ Eri, the Sun, and Procyon, considering the complete spectral domain. 
        The gain in \Nu \ is also important, particularly for the Sun.
         
\begin{figure*}[h!]
        \centering              
        \includegraphics[width=\hsize]{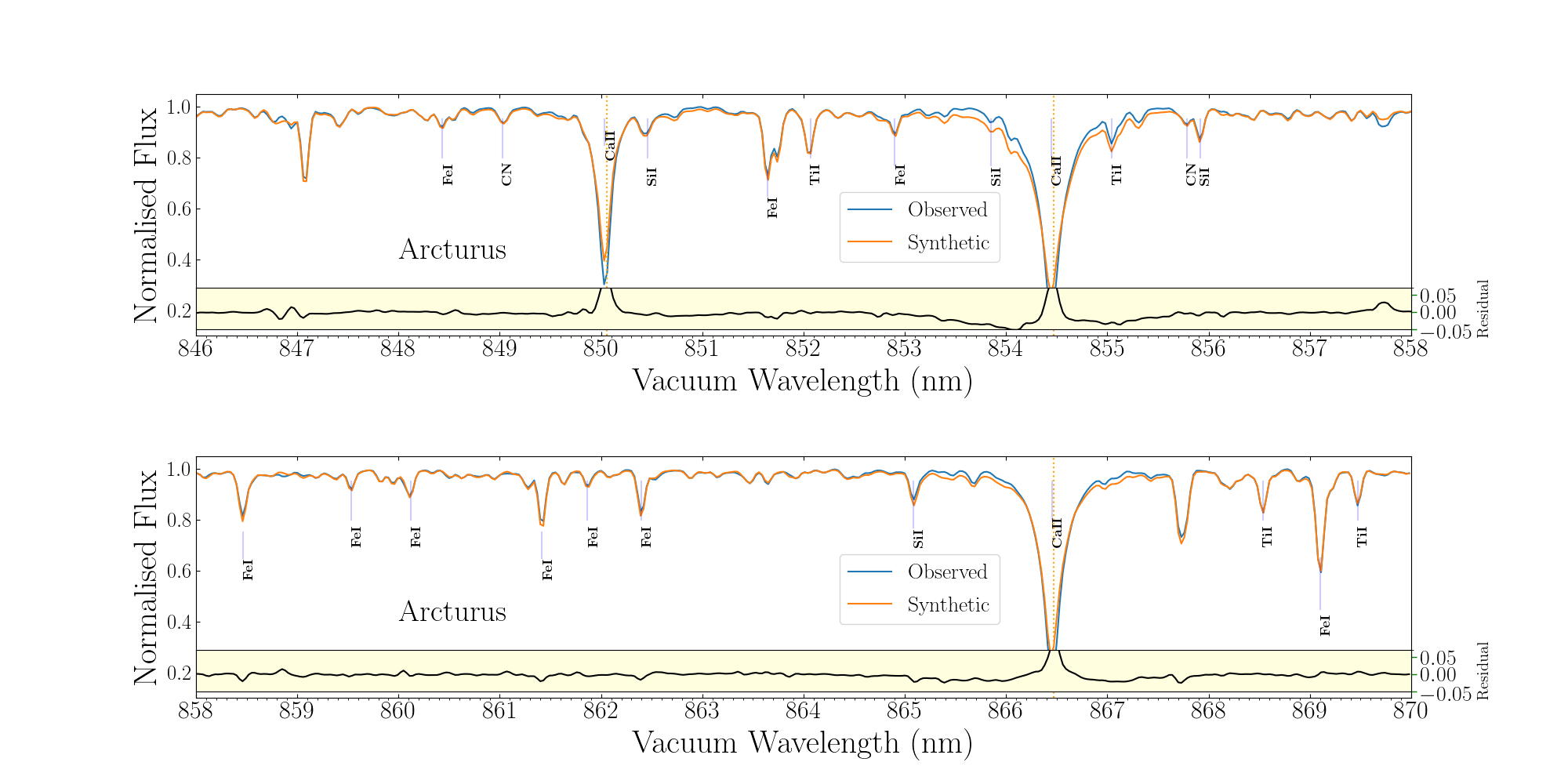} 
\caption{Observed (blue) and synthetic (orange) spectra for Arcturus. Vertical green and orange dashed lines identify spectral lines 
        differing by between 3\% and 5\% and by more than 5\% in relative flux, respectively. These spectra were computed adopting the final GSP-Spec line list and are shown at the RVS spectral resolution with 800 $wlp$. We note that the wavelengths are shown in vacuum. Some line identifications are shown for non-blended absorption lines with a 
        normalised flux below 0.93. The lower yellow insets show the difference between the observed and synthetic fluxes.}
        \label{fig:Arcturus}
\end{figure*}

\begin{figure*}[h!]
        \centering
        \includegraphics[width=\hsize]{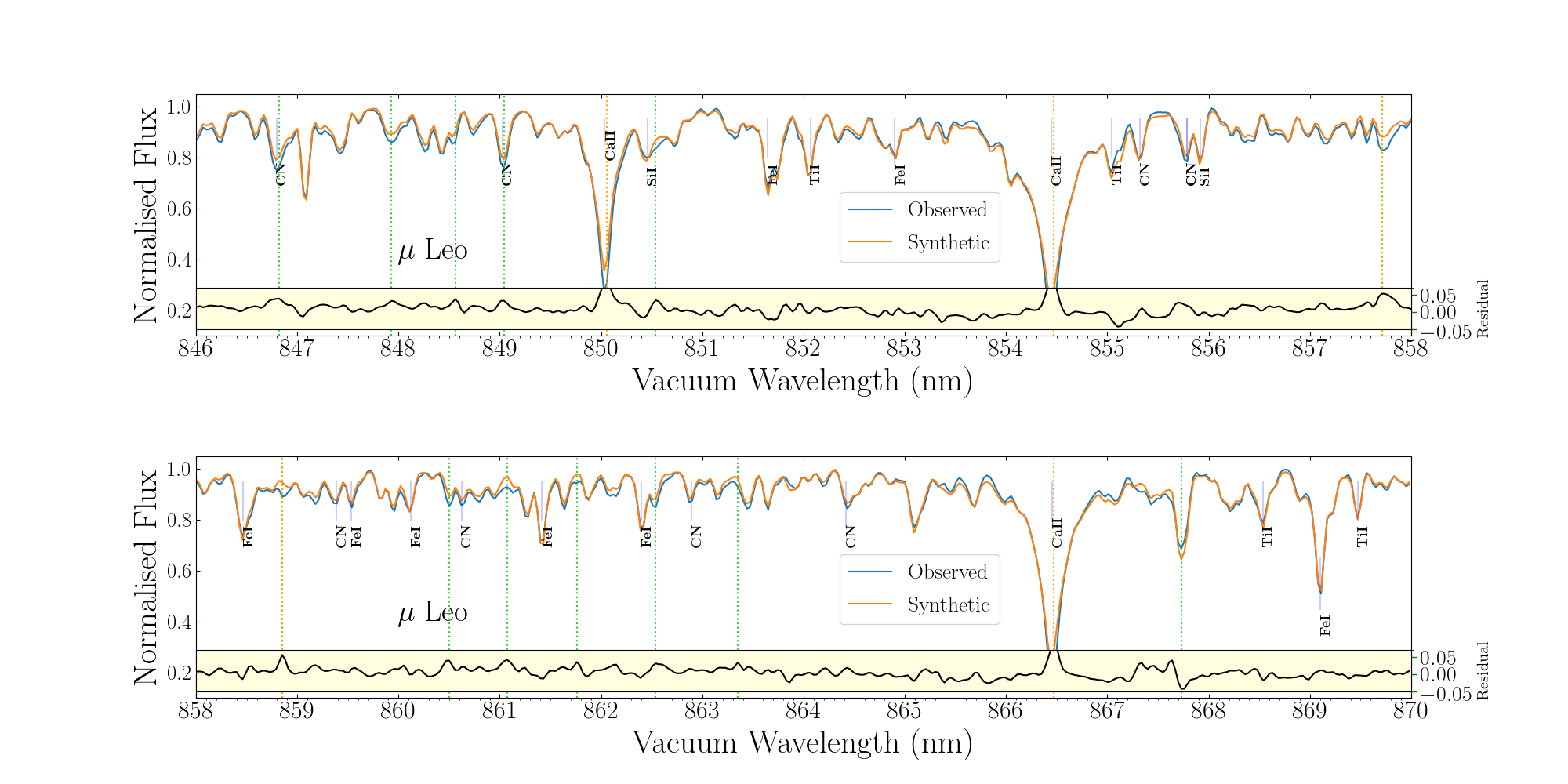} 
        \caption{Same as Fig.~\ref{fig:Arcturus}, but for $\mu$ Leo with an identified line with a depth lower than 0.85.}
                \label{fig:MuLeo}
        \end{figure*}

\begin{figure*}[h!]
        \includegraphics[width=\hsize]{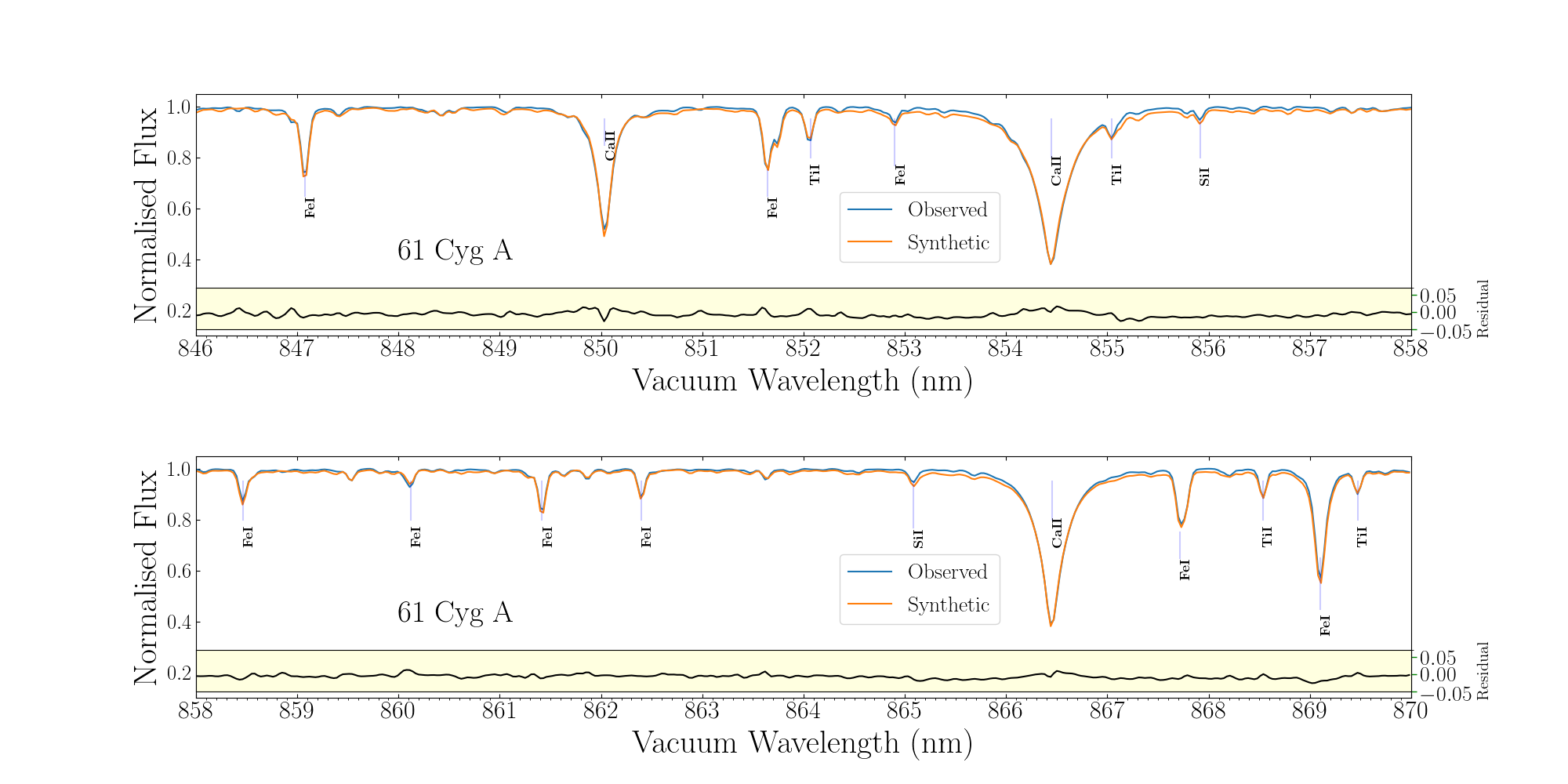}
        \caption{Same as Fig.~\ref{fig:Arcturus}, but for 61~Cyg~A with identified lines that have a depth lower than 0.96.}
        \label{fig:61CygA}
        \end{figure*}

\begin{figure*}[h!]
                \includegraphics[width=\hsize]{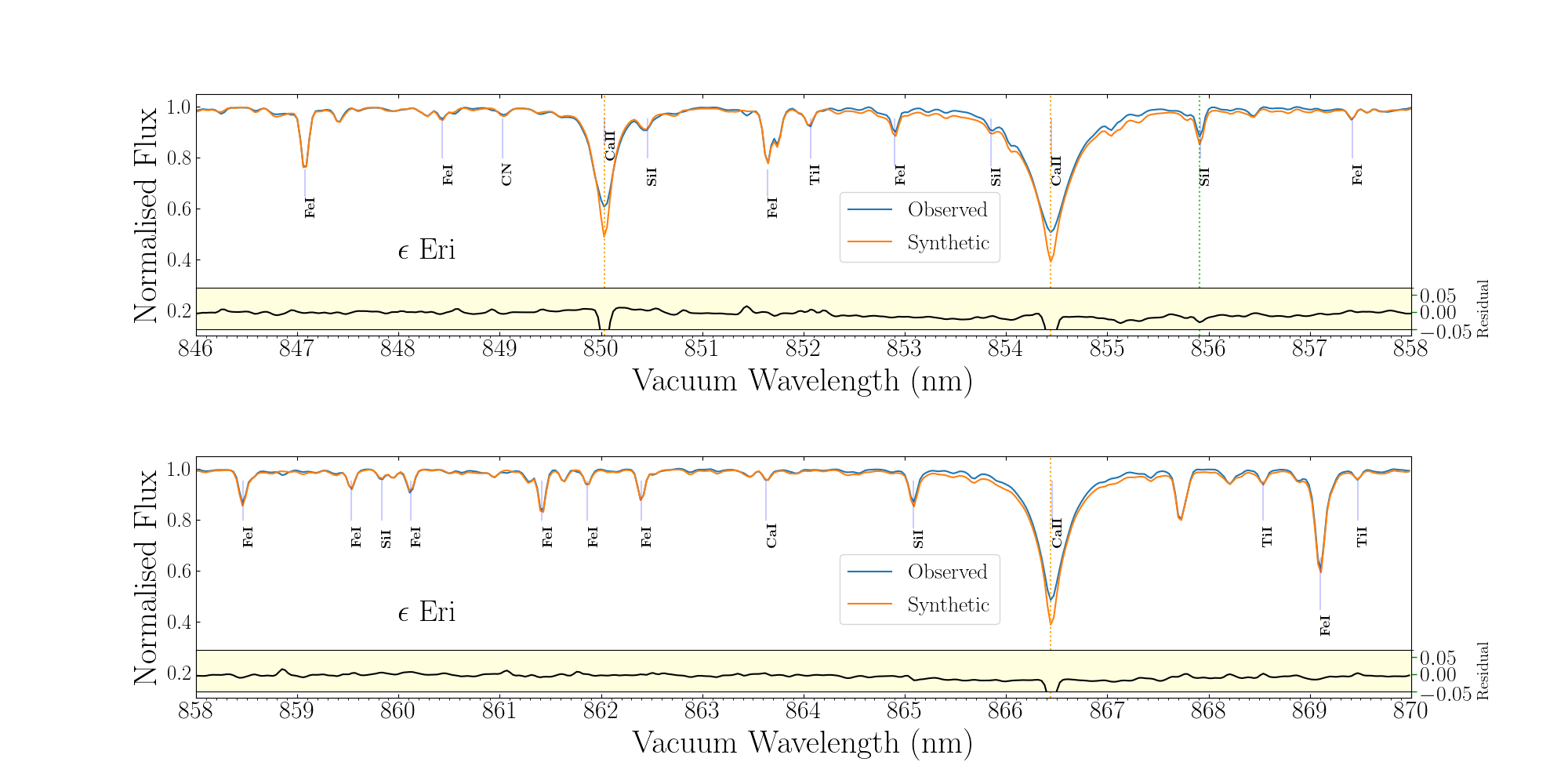} 
                \caption{Same as Fig.~\ref{fig:Arcturus}, but for $\epsilon$~Eri with identified lines that have a depth lower than 0.96.}
                \label{fig:EpsEri}
\end{figure*}

\begin{figure*}[h!]
                \includegraphics[width=\hsize]{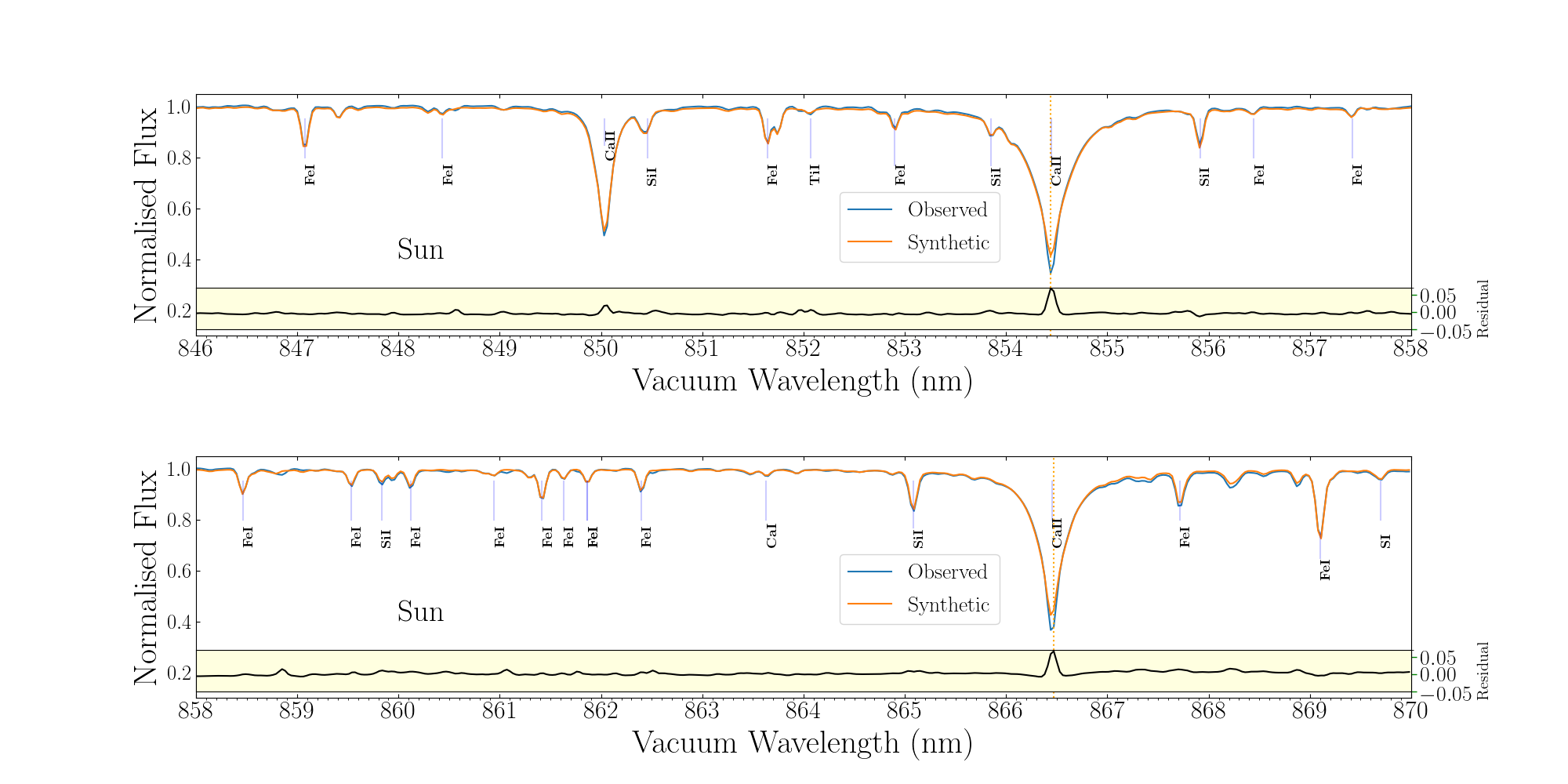} 
                \caption{Same as Fig.~\ref{fig:Arcturus}, but for the Sun, which has identified lines that have a depth lower than 0.97.}
                \label{fig:Sun}
\end{figure*}

\begin{figure*}[h!]
                \includegraphics[width=\hsize]{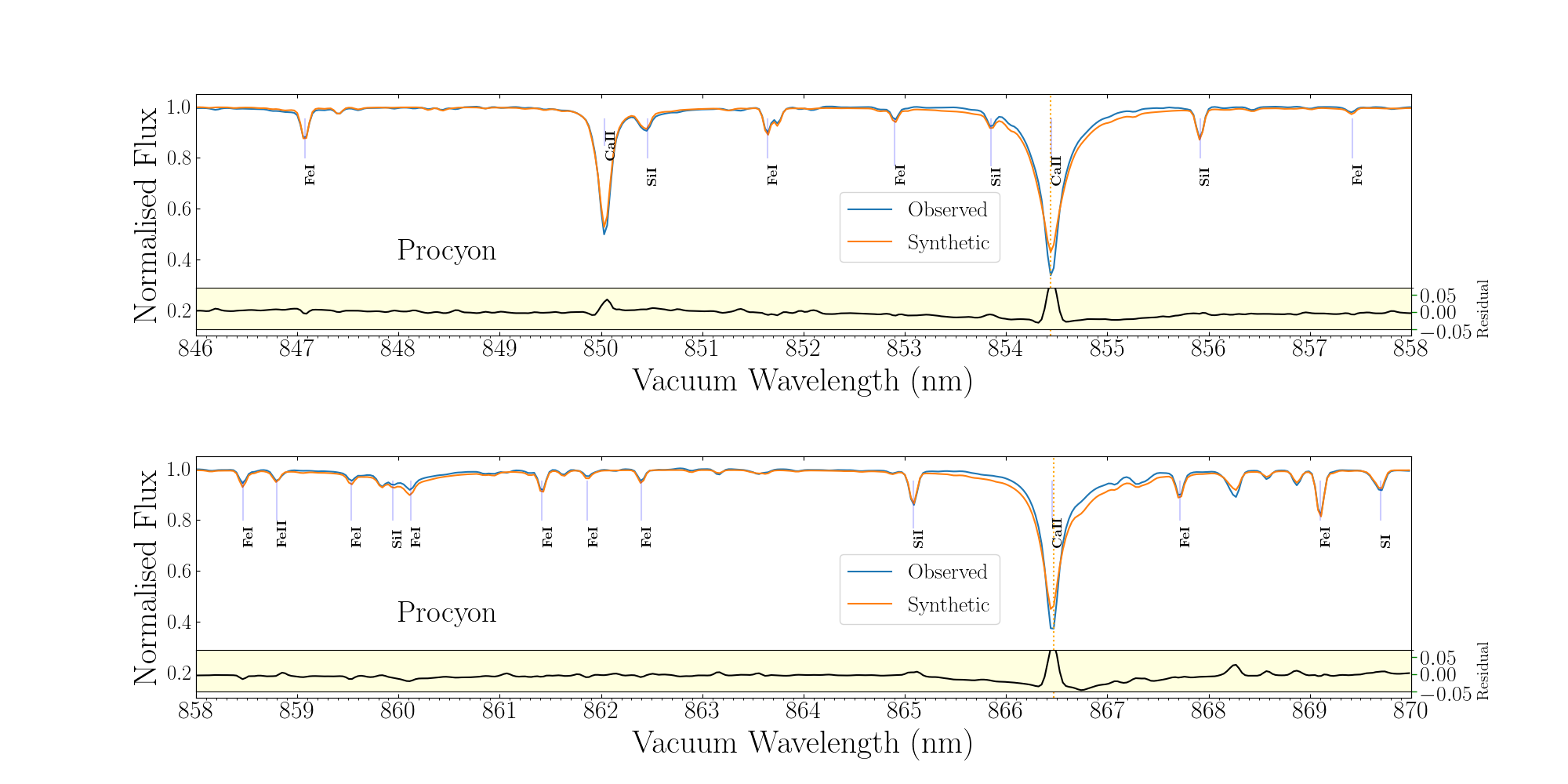} 
        \caption{Same as Fig.~\ref{fig:Arcturus}, but for Procyon, which has identified lines with a depth lower than 0.97.}
        \label{fig:Procyon}
\end{figure*}

        The improved fit quality is also illustrated in Figs.~\ref{fig:Arcturus} to \ref{fig:Procyon}, which
        show the observed (in blue) and synthetic (in orange) spectra for the six reference stars at the RVS spectral resolution (these star global fits are ordered as
        in Tab.~\ref{RefStars}). 
        One can again see an excellent global match between both spectra, confirming
        the good QFP reported in column $G$L of Tab.~\ref{GL_11500}. 
        Visually, the Sun and 61~Cyg~A exhibits the best fits, whereas $\mu$ Leo shows the largest mismatches.
        We can also remark that the largest mismatches (shown by green and red vertical dashed lines corresponding to \Nt~and \Nc, respectively) originate from the calcium triplet lines. Both 
        the cores (for all the stars excepted 61~Cyg~A) and wings (especially for Procyon) are responsible for the main discrepancies. However, we highlight the good fit of the wings of triplet calcium lines for the Sun. 
        
        In summary, if one disregards the calcium triplet lines and some faint 
        molecular lines causing most of the small differences in the coolest stars, 
        the derived GSP-Spec line list strongly improves the 
        fit quality for the studied reference stars.
        This new line list can therefore be safely adopted for the analysis of 
        $Gaia$/RVS spectroscopic data.


\section{Conclusion}
\label{Sect:ccl}
The automatic analysis of $Gaia$-RVS spectra performed by the DPAC/GSP-Spec group
in order to estimate the atmospheric parameters and individual chemical abundances
of late-type stars relies on synthetic spectra grids. 
To compute high-quality spectra for such cool stars, 
good quality line data are required. 
In this article, we thus present the line list that was adopted within GSP-Spec. 

In order to quantify the line list quality, we first selected
six very well-parametrised reference giant and dwarf 
stars with effective temperatures ranging from $\sim$4,200~K to $\sim$6,500~K,
that is, representing the typical \T \ of the stars presently analysed
within GSP-Spec. Observed high-resolution and high-S/N spectra 
covering the RVS spectral domain were retrieved
for these reference stars.
Then, high-resolution synthetic spectra were computed for
these reference stars thanks to the TURBOSPECTRUM v19.1.2 radiative transfer code
and MARCS model atmosphere. 

Stellar synthetic spectra were first computed by adopting
the line list built by the $Gaia$-ESO Survey consortium.
The quality of this GES line list was quantified by comparing
the observed and simulated spectra at high (R = 100,000) and RVS-like spectral
resolution (R = 11,500). A rather good global agreement for the six reference
stars is found, although some important line mismatches are reported.
For instance, relative flux differences larger than 3\% \ are identified
for 162 atomic transitions. Smaller disagreements (typically
of the order of 1\%) are mostly caused by molecular transitions.
We find that most of the atomic line mismatches can be solved by correcting
the adopted atomic line data and that the possible incompleteness of the line list 
has a minor impact. We therefore carefully checked the quality of these
line data and, when necessary, we astrophysically calibrated
several atomic oscillator strengths (a couple of line positions were also corrected) in order to fit the observed reference
star spectra as well as possible. The improvement of the flux fit quality adopting
this GSP-Spec line list was quantified
and is found to be important. This new line list produces still not 
perfect synthetic spectra, however, for every type of star,
since a few lines are found to be sometimes not perfectly well reproduced. 
In particular, the ionised calcium triplet lines exhibit mismatches
in their core and/or wings, probably due to NLTE and/or chromospheric effects 
not considered in the adopted synthesis tools. 

Finally, since the new GSP-Spec line list is found to 
produce better and more realistic synthetic spectra, it was adopted for 
the GSP-Spec analysis performed for the thirst data release of $Gaia$.
This line list can be retrieved by contacting the authors of the present
article. However, we warn that it was optimised in the MARCS models
and TURBOSPECTRUM v19.1.2 contexts and should therefore be used with these
same tools. Finally, we notice that no optimisation of the molecular
transitions were performed, and the fit quality outside the 4,000-8,000~K
effective temperature regimes was not checked. Future extensions
of the present work should first focus on these cooler and hotter stars.
More realistic physical assumptions such as NLTE and 3D hydrodynamical simulations should also be 
investigated. Considering the Stark line broadening may also help to better reproduce
some line profiles such as the calcium triplet lines.



%

\begin{acknowledgements}
        The authors warmly thank Bengt Edvardsson for sharing his fruitfull
        experience for building line list. 
        We are grateful to M.T. Belmonte Sainz-Ezquerra, Y. Frémat, A. Lobel, J. Pickering, and N. Ryde for discussions on some specific atomic transitions.The huge efforts performed by the GES line list group to   produce their lists are also thanked. We acknowledge M. Bergemann for providing this GES line list
        in TURBOSPECTRUM format. We also sincerely thank the stellar atmosphere group in Uppsala
        for providing the MARCS model atmospheres to the community, B. Plez for having 
        developped and maintaining the TURBOSPECTRUM package and, all the atomic/molecular line list
        providers. In particular, B. Plez and T. Masseron are acknowledged for having provided 
        some molecular line lists. This work has made use of the VALD database, operated at
         Uppsala University, the Institute of Astronomy RAS in Moscow, and the University of Vienna.
         We also used the SIMBAD database, operated at CDS, Strasbourg, France.
         Part of the calculations have been performed with the high-performance computing 
         facility SIGAMM, hosted by OCA.
         Finally, we thank the anonymous referee for their valuable comments.
\end{acknowledgements}

\bibliographystyle{aa} 
\bibliography{ref}

\end{document}